\documentclass[preprint,preprintnumbers,amsmath,amssymb]{revtex4}

\usepackage{graphicx}
\usepackage{dcolumn}

\begin{document}


\title{Derivation of Nonlinear Evolution Equations
 for Coupled and Single Fields in a Quadratic Medium}

\author{Jeffrey Moses and Frank W. Wise}
\affiliation{Department of Applied and Engineering Physics, Cornell
University, Ithaca, New York 14853, USA}


\date{\today}

\begin{abstract}
We derive coupled propagation equations for ultrashort pulses in a
degenerate three-wave mixing process in quadratic media, using
approximations consistent with the slowly evolving wave
approximation [T. Brabec and F. Krausz, Phys. Rev. Lett.
\textbf{78}, 3282 (1997)].  From these we derive an approximate
single-field equation for the fundamental field.  This document
expands upon mathematics used for work submitted by the same authors
to Physical Review Letters.
\end{abstract}

\maketitle                   

\section{Introduction}      

The following derivation has two main parts.  First, Maxwell's
equations are reduced to a set of coupled propagation equations for
degenerate three-wave mixing in quadratic media with less
restriction on the minimum pulse duration than required by the
Slowly Varying Envelope Approximation (SVEA). To do this we use a
set of approximations consistent with those used by Brabec and
Krausz in their derivation of the Nonlinear Evolution Equation (NEE)
\cite{brabec}. These are known as the ``Slowly Evolving Wave
Approximation (SEWA)''.  Second, we collapse the coupled propagation
equations to an approximate single-field equation for the
fundamental field using perturbation methods.

\section{Derivation part 1a: generalized equations}

We begin with Maxwell's equations for the electric field vector
$\mathbf{E}$ in the Fourier domain, including linear
$\mathbf{P}^{(1)}$ and quadratic nonlinear $\mathbf{P}^{(2)}$
contributions to the electric polarization,

\begin{equation} \label{eq1}
\left(\frac{\partial^2}{\partial z^2} +
    \nabla^2_\bot \right)\mathbf{E}(\mathbf{r},\omega)
    + \frac{\omega^2}{c^2} \mathbf{E}(\mathbf{r},\omega)
    =
    - \frac{4\pi\omega^2}{c^2}
        [\mathbf{P}^{(1)}(\mathbf{r},\omega)
            + \mathbf{P}^{(2)}(\mathbf{r},\omega)].
\end{equation}

Next, we expand the electric field and polarization into waves
propagating at the fundamental (FF) and second harmonic (SH)
frequencies.  In performing this step we assume the FF and SH
spectra do not overlap.  Here, $\hat{e}_1$ and $\hat{e}_2$ denote
the polarization of the FF and SH fields, respectively.
{\setlength\arraycolsep{2pt}
\begin{eqnarray} \label{eq2}
\mathbf{E}(\mathbf{r},\omega)
    &=& \mathbf{E}_{tot}(\mathbf{r},\omega) = E_1(\mathbf{r},\omega)\hat{e}_1
    + E_2(\mathbf{r},\omega)\hat{e}_2,
\\
\mathbf{P}^{(1)}(\mathbf{r},\omega)
    &=& \chi ^{(1)}_1(\omega)\cdot \mathbf{E}_{tot}(\mathbf{r},\omega)
    + \chi ^{(1)}_2(\omega)\cdot
    \mathbf{E}_{tot}(\mathbf{r},\omega),
\\
\mathbf{P}^{(2)}(\mathbf{r},\omega)
    &=& \mathbf{E}_{tot}(\mathbf{r},\omega)\cdot\chi ^{(2)}_1(\omega)\cdot
        \mathbf{E}_{tot}(\mathbf{r},\omega)
    + \mathbf{E}_{tot}(\mathbf{r},\omega)\cdot\chi ^{(2)}_2(\omega)\cdot
        \mathbf{E}_{tot}(\mathbf{r},\omega).
\end{eqnarray}}
The subscript $j$ in $\chi ^{(k)}_j(\omega)$ denotes the
$k^{th}$-order susceptibility that gives rise to a FF or SH wave
(for $j = 1,2$, respectively).

Defining the linear indices of refraction
$\left[n_{0,j}(\omega)\right]^2=1+4\pi\chi ^{(1)}_j(\omega)$ and
wavevectors $\left[k_j(\omega)\right]^2=
\left[n_{0,j}(\omega)\right]^2\omega^2 / c^2$ in the usual fashion
and rearranging (1), we obtain a set of coupled-field equations,
\begin{equation} \label{eq5}
\left(\frac{\partial^2}{\partial z^2} + \nabla^2_\bot \right)E_1(\mathbf{r},\omega)
    + \left[k_1(\omega)\right]^2 E_1(\mathbf{r},\omega)
    =
    - \frac{4\pi\omega^2}{c^2}
        \mathbf{E}_{tot}(\mathbf{r},\omega)\cdot\chi ^{(2)}_1(\omega)\cdot
            \mathbf{E}_{tot}(\mathbf{r},\omega),
\end{equation}
\begin{equation} \label{eq6}
\left(\frac{\partial^2}{\partial z^2} + \nabla^2_\bot \right) E_2(\mathbf{r},\omega)
    + \left[k_2(\omega)\right]^2 E_2(\mathbf{r},\omega)
    =
    - \frac{4\pi\omega^2}{c^2}
        \mathbf{E}_{tot}(\mathbf{r},\omega)\cdot\chi ^{(2)}_2(\omega)\cdot
            \mathbf{E}_{tot}(\mathbf{r},\omega).
\end{equation}

We now introduce the complex envelopes $A_j$ and their Fourier
transforms, {\setlength\arraycolsep{2pt}
\begin{eqnarray} \label{eq7}
E_1(\mathbf{r},t) &=&
    A_1(\mathbf{r}_\bot,z,t)e^{i(k_1(\omega_0)z - \omega_0t)} + \textrm{c.c.}
    \longleftrightarrow
    A_1(\mathbf{r}_\bot,z,\omega - \omega_0)e^{i(k_1(\omega_0)z)} +
    \textrm{c.c.},
\\
E_2(\mathbf{r},t) &=&
    A_2(\mathbf{r}_\bot,z,t)e^{i(k_2(2\omega_0)z - 2\omega_0t)} + \textrm{c.c.}
    \longleftrightarrow
    A_2(\mathbf{r}_\bot,z,\omega - 2\omega_0)e^{i(k_2(2\omega_0)z)} + \textrm{c.c.}
\end{eqnarray}}

Substituting envelopes (7),(8) into equations (5),(6) and
defining $\Delta k = 2 k_1 - k_2$, we obtain
\begin{eqnarray} \label{eq9}
\left(2 i k_1(\omega_0) \frac{\partial}{\partial z}
        + \frac{\partial^2}{\partial z^2}
        - \left[k_1(\omega_0)\right]^2 + \nabla^2_\bot\right) A_1(\mathbf{r}_\bot,z, \omega - \omega_0)
    \nonumber\\
    + \left[k_1(\omega)\right]^2 A_1(\mathbf{r}_\bot,z, \omega - \omega_0)
    \nonumber\\
    = - \frac{8\pi\omega^2}{c^2}
        \chi ^{(2)}_{2\omega_0 - \omega_0 = \omega_0}
        A_1^\ast(\mathbf{r}_\bot,z, \omega + \omega_0)
        A_2(\mathbf{r}_\bot,z, \omega - 2\omega_0) e^{- i \Delta k
        z},
\\
\left(2 i k_2(2\omega_0) \frac{\partial}{\partial z}
        + \frac{\partial^2}{\partial z^2}
        - \left[k_2(2\omega_0)\right]^2 + \nabla^2_\bot\right) A_2(\mathbf{r}_\bot,z, \omega - 2\omega_0)
    \nonumber\\
    + \left[k_2(\omega)\right]^2 A_2(\mathbf{r}_\bot,z, \omega - 2\omega_0)
    \nonumber\\
    {}= - \frac{4\pi\omega^2}{c^2}
        \chi ^{(2)}_{\omega_0 + \omega_0 = 2\omega_0}
        A_1^2(\mathbf{r}_\bot,z, \omega - \omega_0) e^{i \Delta k
        z},
\end{eqnarray}
(and complex conjugates).  Henceforth, we'll ignore the conjugate
equations.

As usual, we expand wavevectors $k_j(\omega_j)$ about $\omega_j$
($\omega_1 = \omega_0$, $\omega_2 = 2\omega_0$):
\begin{equation} \label{eq11}
k_j(\omega) = k_j(\omega_j) + k'_j(\omega - \omega_j)
    + D_j(\omega - \omega_j),
\end{equation}
with $k_j^n = \frac{d^n k_j}{d \omega^n}|_{\omega_j}$ and
$D_j(\omega -\omega_j)
    = \sum_{n=2}^\infty \frac{k_j^n}{n!} (\omega - \omega_j)^n$.

We will now assume the quadratic nonlinearity is instantaneous, i.e.
non-instantaneous effects are negligible $\left(\frac{d^n
\chi_j^{(2)}}{d\omega^n}|_{\omega_j} \simeq 0\right)$, and perform
the reverse transform:
\begin{eqnarray} \label{eq12}
\left[2 i k_1 \frac{\partial}{\partial z}
    + \frac{\partial^2}{\partial z^2}
    - k_1^2 + \nabla^2_\bot\right] A_1(\mathbf{r}_\bot,z,t)
\nonumber\\
{}+ \left[k_1 + i k'_1 \frac{\partial}{\partial t}
    + D_1(t)\right]^2 A_1(\mathbf{r}_\bot,z,t)
\nonumber\\
= - \frac{8 \pi}{c^2}\left(\omega_0 + i \frac{\partial}{\partial
    t}\right)^2 A_1^\ast(\mathbf{r}_\bot,z,t) A_2(\mathbf{r}_\bot,z,t) e^{- i \Delta k
    z},
\\
\left[2 i k_2 \frac{\partial}{\partial z}
    + \frac{\partial^2}{\partial z^2}
    - k_2^2 + \nabla^2_\bot\right] A_2(\mathbf{r}_\bot,z,t)
\nonumber\\
{}+ \left[k_2 + i k'_2 \frac{\partial}{\partial t}
    + D_2(t)\right]^2 A_2(\mathbf{r}_\bot,z,t)
\nonumber\\
= - \frac{4 \pi}{c^2}\left(2\omega_0 + i \frac{\partial}{\partial
    t}\right)^2 A_1^2(\mathbf{r}_\bot,z,t) e^{i \Delta k z},
\end{eqnarray}
where $D_j(t) = \sum_{n=2}^\infty \frac{k_j^n}{n!}
\left(i\frac{\partial}{\partial t}\right)^n$.  Note: $\omega^2 =
(\omega-\omega_0)^2 + 2(\omega-\omega_0)\omega_0 + \omega_0^2$, and
thus $\mathcal{F}^{-1}[\omega^2 G(\omega)] = -\frac{\partial^2
G(t)}{\partial t^2} + 2 i \omega_0 \frac{\partial G(t)}{\partial t}
+ \omega_0^2 = (\omega_0 + i \frac{\partial}{\partial t})^2 G(t)$.
Also, $\omega^2 = (\omega-2\omega_0)^2 +
2(\omega-2\omega_0)(2\omega_0) + (2\omega_0)^2$.

Next, we transform coordinate systems to co-moving variables in the
frame of the FF wave, $\tau = t - k'_1 z$, $z' = z$.  Partial
derivatives become
{\setlength\arraycolsep{2pt}
\begin{eqnarray} \label{eq14}
\frac{\partial}{\partial t} &=& \frac{\partial}{\partial \tau},
\\
\frac{\partial^2}{\partial t^2} &=& \frac{\partial^2}{\partial
\tau^2},
\\
\frac{\partial}{\partial z} &=& \frac{\partial}{\partial z'}
    - k'_1 \frac{\partial}{\partial \tau},
\\
\frac{\partial^2}{\partial z^2} &=& \frac{\partial^2}{\partial z'^2}
    - 2 k'_1 \frac{\partial^2}{\partial z' \tau}
    + k_1^{\prime2} \frac{\partial^2}{\partial \tau^2}.
\end{eqnarray}}
(We drop the prime on $z$ for notational simplicity) and the
equation for the FF becomes
\begin{eqnarray} \label{eq18}
\left[2 i k_1 \left(\frac{\partial}{\partial z}
        - k'_1\frac{\partial}{\partial\tau}\right)
    + \left(\frac{\partial^2}{\partial z^2}
        - 2 k'_1\frac{\partial^2}{\partial z \partial \tau}
        + k_1^{\prime2}\frac{\partial^2}{\partial \tau^2}\right)
    - k_1^2 + \nabla^2_\bot\right] A_1
\nonumber\\
{}+ \left[k_1^2
    + 2 i k_1 k'_1 \frac{\partial}{\partial \tau}
    + 2 k_1 D_1
    - k_1^{\prime2} \frac{\partial^2}{\partial \tau^2}
    + 2 i k'_1 D_1 \frac{\partial}{\partial \tau}
    + D_1^2\right] A_1
\nonumber\\
= - \frac{8 \pi}{c^2}\left(\omega_0 + i \frac{\partial}{\partial
    \tau}\right)^2 \chi ^{(2)}_{2\omega_0 - \omega_0}
         A_1^\ast A_2 e^{- i \Delta k z}.
\end{eqnarray}

Combining terms, the equation for the FF simplifies to
\begin{eqnarray} \label{eq19}
2 i k_1 \left(1 + i \frac{k'_1}{k_1}
    \frac{\partial}{\partial\tau}\right) \frac{\partial A_1}{\partial z}
= - \nabla^2_\bot A_1 - \frac{\partial^2 A_1}{\partial z^2}
    - 2 k_1 D_1 \left(1 + i \frac{k'_1}{k_1}
        \frac{\partial}{\partial\tau}\right) A_1
    -D_1^2 A_1
\nonumber\\
- \frac{8 \pi \omega_0^2 \chi ^{(2)}_{2\omega_0 -
    \omega_0}}{c^2}\left(1 + \frac{i}{\omega_0} \frac{\partial}{\partial
    \tau}\right)^2 A_1^\ast A_2 e^{- i \Delta k z}.
\end{eqnarray}

The equation for the SH becomes
\begin{eqnarray} \label{eq20}
\left[2 i k_2 \left(\frac{\partial}{\partial z}
        - k'_1\frac{\partial}{\partial\tau}\right)
    + \left(\frac{\partial^2}{\partial z^2}
        - 2 k'_1\frac{\partial^2}{\partial z \partial \tau}
        + k_1^{\prime2}\frac{\partial^2}{\partial \tau^2}\right)
    - k_2^2 + \nabla^2_\bot\right] A_2
\nonumber\\
{}+ \left[k_2^2
    + 2 i k_2 k'_2 \frac{\partial}{\partial \tau}
    + 2 k_2 D_2
    - k_2^{\prime2} \frac{\partial^2}{\partial \tau^2}
    + 2 i k'_2 D_2 \frac{\partial}{\partial \tau}
    + D_2^2\right] A_2
\nonumber\\
= - \frac{4 \pi}{c^2}\left(2\omega_0 + i \frac{\partial}{\partial
    \tau}\right)^2 \chi ^{(2)}_{\omega_0 + \omega_0}
         A_1^2 e^{i \Delta k z}.
\end{eqnarray}

Since our coordinate system is in the frame of the FF wave, fewer
terms will cancel when we combine terms in the SH equation:
\begin{eqnarray} \label{eq21}
2 i k_2 \left(1 + i \frac{k'_1}{k_2}
    \frac{\partial}{\partial\tau}\right) \frac{\partial A_2}{\partial z}
= - \nabla^2_\bot A_2 - \frac{\partial^2 A_2}{\partial z^2}
    - 2 k_2 D_2 \left(1 + i \frac{k'_2}{k_2}
        \frac{\partial}{\partial\tau}\right) A_2
    -D_2^2 A_2
\nonumber\\
- 2 i k_2 (k'_2 - k'_1) \frac{\partial A_2}{\partial \tau}
    -(k_1^{\prime2} - k_2^{\prime2})\frac{\partial^2 A_2}{\partial \tau^2}
\nonumber\\
- \frac{16 \pi \omega_0^2 \chi ^{(2)}_{\omega_0 +
    \omega_0}}{c^2}\left(1 + \frac{i}{2\omega_0} \frac{\partial}{\partial
    \tau}\right)^2 A_1^2 e^{i \Delta k z}.
\end{eqnarray}
The fifth term on the RHS is the familiar group velocity mismatch
(GVM) term, while the sixth term on the RHS is unfamiliar. Also
notable, unlike in the FF equation, the prefactor to the
$\frac{\partial A}{\partial z}$ term has mixed wavevectors by field
$\left(\frac{k'_1}{k_2}\right)$.  Also, the GVM term has no
prefactor $\left(1 + \big(\quad\big) \frac{\partial}{\partial \tau}
\right)$.

Assuming we'll want to use the approximation $v_p \simeq v_g$
$\Big(\frac{1}{\omega_0} \simeq \frac{k'_1}{k_1}$ for FF, and
$\frac{1}{2\omega_0} \simeq \frac{k'_2}{k_2}$ for SH$\Big)$ to
simplify the equations in accordance with Brabec and Krausz's SEWA,
we use some algebra to rearrange terms in the SH equation:
\begin{eqnarray} \label{eq22}
k_1^{\prime2} - k_2^{\prime2}
    &=& (k'_1 - k'_2)(k'_1 + k'_2)
\nonumber\\
    &=& (k'_1 - k'_2)(k'_1 - k'_2 + 2 k'_2)
\nonumber\\
    &=& (k'_1 - k'_2)^2 + 2 k'_2(k'_1 - k'_2).
\end{eqnarray}
Thus, {\setlength\arraycolsep{2pt}
\begin{eqnarray} \label{eq23}
- 2 i k_2 (k'_2 &-& k'_1) \frac{\partial A_2}{\partial \tau}
    -(k_1^{\prime2} - k_2^{\prime2})\frac{\partial^2 A_2}{\partial \tau^2}
\nonumber\\
&=& - 2 i k_2 (k'_2 - k'_1) \frac{\partial A_2}{\partial \tau}
    + 2k'_2(k'_2 - k'_1)\frac{\partial^2 A_2}{\partial \tau^2}
    - (k'_2 - k'_1)^2\frac{\partial^2 A_2}{\partial \tau^2}
\nonumber\\
&=& - 2 i k_2 (k'_2 - k'_1)
        \left(1 + \frac{k'_2}{k_2}\frac{\partial}{\partial \tau}\right)
        \frac{\partial A_2}{\partial \tau}
    - (k'_2 - k'_1)^2\frac{\partial^2 A_2}{\partial \tau^2}.
\end{eqnarray}}
Also, {\setlength\arraycolsep{2pt}
\begin{eqnarray} \label{eq24}
\left(1 + i \frac{k'_1}{k_2} \frac{\partial}{\partial \tau}\right)
    \frac{\partial A_2}{\partial z}
&=&\left(1 + i \frac{(k'_2 - k'_2 + k'_1)}{k_2}
        \frac{\partial}{\partial \tau}\right)
    \frac{\partial A_2}{\partial z}
\nonumber\\
&=&\left(1 + i \frac{k'_2}{k_2}
            \frac{\partial}{\partial \tau}
        - i \frac{(k'_2 - k'_1)}{k_2}
            \frac{\partial}{\partial \tau}\right)
    \frac{\partial A_2}{\partial z}
\nonumber\\
&=&\left(1 + i \frac{k'_2}{k_2}
            \frac{\partial}{\partial \tau}\right)
        \frac{\partial A_2}{\partial z}
    - i \frac{(k'_2 - k'_1)}{k_2}
        \frac{\partial^2 A_2}{\partial z \partial \tau}.
\end{eqnarray}}

Using (23) and (24), the SH equation becomes
{\setlength\arraycolsep{2pt}
\begin{eqnarray} \label{eq25}
2 i k_2 \left(1 + i \frac{k'_2}{k_2}
    \frac{\partial}{\partial\tau}\right) \frac{\partial A_2}{\partial z}
&=& - \nabla^2_\bot A_2 - \frac{\partial^2 A_2}{\partial z^2}
    - 2(k'_2 - k'_1)\frac{\partial^2 A_2}{\partial z \partial \tau}
    - (k'_2 - k'_1)^2\frac{\partial^2 A_2}{\partial \tau^2}
\nonumber\\
& & {}- 2 i k_2 (k'_2 - k'_1) \left(1 + i \frac{k'_2}{k_2}
    \frac{\partial}{\partial\tau}\right)\frac{\partial A_2}{\partial \tau}
\nonumber\\
& & {} - 2 k_2 D_2 \left(1 + i \frac{k'_2}{k_2}
        \frac{\partial}{\partial\tau}\right) A_2
    -D_2^2 A_2
\nonumber\\
& & {}- \frac{16 \pi \omega_0^2 \chi ^{(2)}_{\omega_0 +
    \omega_0}}{c^2}\left(1 + \frac{i}{2\omega_0} \frac{\partial}{\partial
    \tau}\right)^2 A_1^2 e^{i \Delta k z},
\end{eqnarray}}
where the unfamiliar third and fourth terms on the RHS are due to
our choice of co-moving variables in the frame of the FF wave.

Dividing through by the prefactor to the $\frac{\partial A}{\partial
z}$ term in each equation, our two generalized propagation equations
for FF and SH waves with instantaneous quadratic nonlinearity are
(with $\delta = k'_1 - k'_2$)
\begin{eqnarray} \label{eq26}
i\frac{\partial A_1}{\partial z}
    + \frac{1}{2 k_1} \left(1 + i \frac{k'_1}{k_1}
    \frac{\partial}{\partial\tau}\right)^{-1}
        \nabla^2_\bot A_1
    + D_1 A_1
\nonumber\\
    + \frac{4 \pi \omega_0^2
    \chi ^{(2)}_{2\omega_0 - \omega_0}}{k_1 c^2}
    \frac{\left(1 + \frac{i}{\omega_0} \frac{\partial}{\partial
    \tau}\right)^2}{\left(1 + i \frac{k'_1}{k_1}
    \frac{\partial}{\partial\tau}\right)}
    A_1^\ast A_2 e^{- i \Delta k z} =
\nonumber\\
- \frac{1}{2 k_1} \left(1 + i \frac{k'_1}{k_1}
    \frac{\partial}{\partial\tau}\right)^{-1}
    \left[\frac{\partial^2}{\partial z^2} + D_1^2 \right] A_1,
\end{eqnarray}
\begin{eqnarray} \label{eq27}
i\frac{\partial A_2}{\partial z}
    + \frac{1}{2 k_2} \left(1 + i \frac{k'_2}{k_2}
    \frac{\partial}{\partial\tau}\right)^{-1}
        \nabla^2_\bot A_2
    - i\delta\frac{\partial A_2}{\partial \tau}
    + D_2 A_2
\nonumber\\
+ \frac{8 \pi \omega_0^2
    \chi ^{(2)}_{\omega_0 + \omega_0}}{k_2 c^2}
    \frac{\left(1 + \frac{i}{2\omega_0} \frac{\partial}{\partial
    \tau}\right)^2}{\left(1 + i \frac{k'_2}{k_2}
    \frac{\partial}{\partial\tau}\right)}
    A_1^2 e^{i \Delta k z} =
\nonumber\\
- \frac{1}{2 k_2} \left(1 + i \frac{k'_2}{k_2}
    \frac{\partial}{\partial\tau}\right)^{-1}
    \left[\frac{\partial^2}{\partial z^2} + D_2^2
        + 2\delta\frac{\partial^2}{\partial z \partial \tau}
        + \delta^2\frac{\partial^2}{\partial \tau^2} \right] A_2.
\end{eqnarray}

\section{Derivation Part 1b: The Slowly Evolving Wave Approximation}

Here we make the first approximations (other than that of
instantaneous quadratic nonlinearity) that place a restriction on
the minimum-duration pulses our equations will accurately model.
Brabec and Krausz point out that the conditions of the SVEA,
\begin{eqnarray} \label{eq28}
\left|\frac{\partial A_j}{\partial z}\right| \ll k_j|A_j|,
\\
\left|\frac{\partial A_j}{\partial \tau}\right| \ll \omega_j|A_j|,
\end{eqnarray}
can be relaxed by replacing (29) with
\begin{equation} \label{eq30}
\left|\frac{k_j - \omega_j k'_j}{k_j}\right| \ll 1,
\end{equation}
(i.e. phase and group velocities are approximately equal), resulting
in the NEE.  Brabec and Krausz called this combination of
approximations, (28), (29), the SEWA, since it can be written
compactly as
\begin{equation} \label{eq31}
\left|\frac{\partial E_j}{\partial z}\right| \ll k_j|E_j|.
\end{equation}
Equation (31) illuminates a key difference between the SVEA and
SEWA: the SEWA requires the \emph{field} rather than \emph{envelope}
to not change significantly as it propagates over the length scale
of a wavelength, and the approximation no longer directly restricts
the bandwidth of the wave to be smaller than its carrier frequency
(or the pulse duration to be much longer than a single optical
cycle) (29).

Condition (30) will be satisfied for our coupled FF and SH waves if
both $\left(\frac{1}{\omega_0} \simeq \frac{k'_1}{k_1}\right)$ and
$\left(\frac{1}{2\omega_0} \simeq \frac{k'_2}{k_2}\right)$ are true.
Applying these conditions to equations (26) and (27), the equations
reduce to
\begin{eqnarray} \label{eq32}
i\frac{\partial A_1}{\partial z}
    + \frac{1}{2 k_1} \left(1 + \frac{i}{\omega_0}
    \frac{\partial}{\partial\tau}\right)^{-1}
        \nabla^2_\bot A_1
    + D_1 A_1
    \nonumber\\
    + \frac{4 \pi \omega_0^2
    \chi ^{(2)}_{2\omega_0 - \omega_0}}{k_1 c^2}
    \left(1 + \frac{i}{\omega_0} \frac{\partial}{\partial
    \tau}\right) A_1^\ast A_2 e^{- i \Delta k z} = 0,
\end{eqnarray}
\begin{eqnarray} \label{eq33}
i\frac{\partial A_2}{\partial z}
    + \frac{1}{2 k_2} \left(1 + \frac{i}{2\omega_0}
    \frac{\partial}{\partial\tau}\right)^{-1}
        \nabla^2_\bot A_2
    - i\delta\frac{\partial A_2}{\partial \tau}
    + D_2 A_2
\nonumber\\
+ \frac{8 \pi \omega_0^2
    \chi ^{(2)}_{\omega_0 + \omega_0}}{k_2 c^2}
    \left(1 + \frac{i}{2\omega_0} \frac{\partial}{\partial
    \tau}\right) A_1^2 e^{i \Delta k z}
\nonumber\\
    = - \frac{1}{2 k_2} \left(1 + \frac{i}{2\omega_0}
    \frac{\partial}{\partial\tau}\right)^{-1}
    \left[2\delta\frac{\partial^2}{\partial z \partial \tau}
        + \delta^2\frac{\partial^2}{\partial \tau^2} \right] A_2,
\end{eqnarray}
where the $\frac{\partial^2 A_j}{\partial z^2}$ terms are negligible
by condition (28), since it also implies
$\frac{1}{k_j}\left|\frac{\partial^2 A_j}{\partial z^2}\right| \ll
\left|\frac{\partial A_j}{\partial z}\right|^\ast$.  We also have
eliminated the $D_j^2 A_j$ terms, assuming all dispersion terms of
order $\frac{\partial^4}{\partial \tau^4}$ and higher will be
negligible.  ($^\ast$ This is a necessary assumption of NEE, NLSE
and all first-order envelope equation derivations, and we similarly
use it here.)

The equation for the FF closely resembles a 1D temporal version of
Brabec and Krausz's NEE, with quadratic nonlinear term replacing
cubic.  However, the SH equation contains the two new terms
\begin{eqnarray} \nonumber
\frac{\delta}{k_2}\frac{\partial^2}{\partial z \partial \tau}
\textrm{ and } \frac{\delta^2}{2 k_2}\frac{\partial^2}{\partial
\tau^2}.
\end{eqnarray}
Our next task is to consider if either of these terms must be
eliminated on the basis of being self-consistent with the
approximations we've already made.

Using a similar approach as P. Kinsler uses in his derivation of the
GFEA equation \cite{kinsler}, we find it is self-consistent at this
point to eliminate the $\frac{\delta}{k_2}\frac{\partial^2}{\partial
z
\partial \tau}$ term, as demonstrated below.
We can write each of our coupled equations in the form
\begin{equation} \label{eq34}
i\frac{\partial A_j}{\partial z} + [LHS_j] = [RHS_j].
\end{equation}
For this equation, the condition
\begin{equation} \label{eq35}
|RHS_j| \ll \left|\frac{\partial A_j}{\partial z}\right|
\end{equation}
is true if and only if
\begin{equation} \label{eq36}
|LHS_j| \simeq \left|\frac{\partial A_j}{\partial z}\right|
\end{equation}
is true as well.  (Note, this simply amounts to $|LHS| \gg
|RHS|$.)  We have already made the approximation (28):
\begin{eqnarray}
\left|\frac{\partial A_j}{\partial z}\right| \ll k_j|A_j|. \nonumber
\end{eqnarray}
Combining (28) and (36), the following condition must hold as well
\begin{equation} \label{eq37}
|LHS_j| \ll k_j|A_j|.
\end{equation}

We have already eliminated $\frac{\partial^2 A_j}{\partial z^2}$
terms by adding a $\frac{\partial}{\partial z}$ to each side of
(28).  Thus, we have assumed (following (37)),
\begin{equation} \label{eq38}
\frac{\partial}{\partial z}|LHS_j| \ll \left|k_j \frac{\partial
A_j}{\partial z}\right|.
\end{equation}
Since
\begin{equation} \label{eq39}
\left|\delta\frac{\partial A_2}{\partial \tau}\right| \leq |LHS_2|,
\end{equation}
it follows that
\begin{equation} \label{eq40}
\left|\frac{\delta}{k_2}\frac{\partial^2 A_2}{\partial z \partial
\tau}\right| \ll \left|\frac{\partial A_2}{\partial z}\right|,
\end{equation}
proving that the elimination of the $\frac{\partial^2 A_j}{\partial
z
\partial \tau}$ terms is consistent with the elimination of
$\frac{\partial^2 A_j}{\partial z^2}$.

However, using the same type of argument we cannot determine that
\begin{equation} \label{eq41}
\left|\frac{\delta^2}{k_2}\frac{\partial^2 A_2}{\partial
\tau^2}\right| \ll \left|\frac{\partial A_2}{\partial z}\right|,
\end{equation}
and it is not consistent to eliminate the
$\frac{\delta^2}{k_2}\frac{\partial^2 A_2}{\partial \tau^2}$ term.

\section{The Nonlinear Evolution Equations in Quadratic Media}

Thus, after using the SEWA our propagation equations become
\begin{eqnarray} \label{eq42}
i\frac{\partial A_1}{\partial z}
    + \frac{1}{2 k_1} \left(1 + \frac{i}{\omega_0}
    \frac{\partial}{\partial\tau}\right)^{-1}
        \nabla^2_\bot A_1
    + D_1 A_1
    \nonumber\\
    + \frac{4 \pi \omega_0^2
    \chi ^{(2)}_{2\omega_0 - \omega_0}}{k_1 c^2}
    \left(1 + \frac{i}{\omega_0} \frac{\partial}{\partial
    \tau}\right) A_1^\ast A_2 e^{- i \Delta k z} = 0,
\end{eqnarray}
\begin{eqnarray} \label{eq43}
i\frac{\partial A_2}{\partial z}
    &+& \frac{1}{2 k_2} \left(1 + \frac{i}{2\omega_0}
    \frac{\partial}{\partial\tau}\right)^{-1}
        \left[\nabla^2_\bot A_2 +
            \delta^2\frac{\partial^2 A_2}{\partial \tau^2}\right]
\nonumber\\
    &-& i\delta\frac{\partial A_2}{\partial \tau}
    + D_2 A_2 + \frac{8 \pi \omega_0^2
    \chi ^{(2)}_{\omega_0 + \omega_0}}{k_2 c^2}
    \left(1 + \frac{i}{2\omega_0} \frac{\partial}{\partial
    \tau}\right) A_1^2 e^{i \Delta k z} = 0,
\end{eqnarray}
which are the Nonlinear Evolution Equations for coupled FF and SH in
a quadratic medium.

After transformation to a nondimensional coordinate system, $\xi =
(\delta/\tau_0) z = z/L_{GVM}$, $s = \tau/\tau_0$, where $\tau_0$ is
the initial temporal half-width, the NEEs become
\begin{eqnarray} \label{eq44}
i\frac{\partial a_1}{\partial \xi}
    + \frac{\rho_1}{2} \left(1 + \frac{i}{\omega_0\tau_0}
    \frac{\partial}{\partial s}\right)^{-1}
        \nabla^2_\bot a_1
    + \mathcal{D}_1 a_1 +
    \left(1 + \frac{i}{\omega_0\tau_0} \frac{\partial}{\partial s}
    \right) a_1^\ast a_2 e^{- i \beta \xi} = 0,
\end{eqnarray}
\begin{eqnarray} \label{eq45}
i\frac{\partial a_2}{\partial \xi}
    + \frac{\rho_2}{2} \left(1 + \frac{i}{2\omega_0\tau_0}
    \frac{\partial}{\partial s}\right)^{-1}
        \left[\nabla^2_\bot a_2 +
            \frac{\nu}{\rho_2}\frac{\partial^2 a_2}{\partial s^2}\right]
    - i\frac{\partial a_2}{\partial s}
    + \mathcal{D}_2 a_2
    \nonumber\\
    + \left(1 + \frac{i}{2\omega_0\tau_0} \frac{\partial}{\partial s}
    \right) a_1^2 e^{i \beta \xi} = 0,
\end{eqnarray}
with definitions,
\begin{eqnarray} \label{eq46}
&&A_i = c_i a_i,\ c_1 = \frac{\delta (k_1 k_2)^{1/2} c^2}
        {4\sqrt{2}\pi\omega_0^2 \tau_0 (\chi ^{(2)}_{2\omega_0 - \omega_0}
            \chi ^{(2)}_{\omega_0 + \omega_0})^{1/2}},\
c_2 = \frac{\delta k_1 c^2}
        {4\pi\omega_0^2 \tau_0 \chi ^{(2)}_{2\omega_0 - \omega_0}},
\nonumber\\
&&\beta = L_{GVM}\Delta k = \frac{\Delta k \tau_0}{\delta},\ \rho_i
=\frac{L_{GVM}}{k_i} = \frac{\tau_0}{\delta k_i},\ \nu =
\frac{L_{GVM} \delta^2}{\tau_0^2 k_2} = \frac{\delta}{\tau_0 k_2},
\nonumber\\
&&\mathcal{D}_i =
\frac{1}{\delta}\sum_{n=2}^{\infty}\frac{k_i^n}{\tau_0^{n-1} n!}
    \left(i\frac{\partial}{\partial s}\right)^n.
\end{eqnarray}

\section{Derivation Part 2: The single-field equation}

Using the same perturbation-method technique as Menyuk \emph{et.
al.} \cite{menyuk}, we collapse coupled propagation equations (44),
(45) to an approximate single-field propagation equation for the
fundamental field (FF).

We employ the transformations $\hat{a}_1 = a_1/|\beta|^{1/2}$ and
$\hat{a}_2 = a_2 e^{-i \beta \xi}$, and restrict our analysis of
transverse dimensions to the temporal only. Furthermore, since
dispersion terms above second order in (44), (45) will not give rise
to significant terms in our derived single-field equation, we will
ignore them here for simplicity. Doing this, we obtain
\begin{eqnarray} \label{eq47}
i\frac{\partial \hat{a}_1}{\partial \xi}
    - \frac{\alpha_1}{2} \frac{\partial^2 \hat{a}_1}{\partial s^2}
    + \left(1 + \frac{i}{\omega_0\tau_0} \frac{\partial}{\partial
        s}\right) (\hat{a}_1^\ast \hat{a}_2)
    = 0,
\end{eqnarray}
\begin{eqnarray} \label{eq48}
i\frac{\partial \hat{a}_2}{\partial \xi} - \beta \hat{a}_2
    - i\frac{\partial \hat{a}_2}{\partial s}
    - \frac{\alpha_2}{2} \frac{\partial^2 \hat{a}_2}{\partial s^2}
    + \frac{\nu}{2} \left(1 + \frac{i}{2\omega_0\tau_0}
    \frac{\partial}{\partial s}\right)^{-1}
        \frac{\partial^2 a_2}{\partial s^2}
    \nonumber\\
    + |\beta|\left(1 + \frac{i}{2\omega_0\tau_0} \frac{\partial}{\partial
        s}\right) (\hat{a}_1^2)
    = 0,
\end{eqnarray}
where $\alpha_j = L_{GVM}/L_{GVD_j} =
k^{\prime\prime}_j/\tau_0\delta$.  We can treat the fifth term of
equation (48) as a perturbation to GVD and higher orders of
dispersion, since they are of the same order in our expansion. Thus,
we define $\alpha'_2 = \alpha_2 - \nu$, and equation (48) becomes
\begin{eqnarray} \label{eq49}
i\frac{\partial \hat{a}_2}{\partial \xi} - \beta \hat{a}_2
    - i\frac{\partial \hat{a}_2}{\partial s}
    - \frac{\alpha'_2}{2} \frac{\partial^2 \hat{a}_2}{\partial s^2}
    + |\beta|\left(1 + \frac{i}{2\omega_0\tau_0} \frac{\partial}{\partial
        s}\right) (\hat{a}_1^2)
    = 0.
\end{eqnarray}

In order to find an approximate expression for $a_2$ in terms of
$a_1$, we now expand the SH field in a power series in $1/|\beta|$,
\begin{equation} \label{eq50}
\hat{a}_2 = \sum_{n=0}^\infty \frac{\hat{a}_{2,n}}{|\beta|^n},
\end{equation}
and substitute this in the coupled equations (47), (49):
\begin{eqnarray} \label{eq51}
i\frac{\partial \hat{a}_1}{\partial \xi}
    - \frac{\alpha_1}{2} \frac{\partial^2 \hat{a}_1}{\partial s^2}
    + \left(1 + \frac{i}{\omega_0\tau_0} \frac{\partial}{\partial
        s}\right) \left(\hat{a}_1^\ast \left[\hat{a}_{20} +
            \frac{\hat{a}_{21}}{|\beta|} + \frac{\hat{a}_{22}}{|\beta|^2} +
            \cdots \right]\right)
    = 0,
\end{eqnarray}
{\setlength\arraycolsep{0pt}
\begin{eqnarray} \label{eq52}
&i&\frac{\partial}{\partial \xi}\left[\hat{a}_{20} +
        \frac{\hat{a}_{21}}{|\beta|} + \frac{\hat{a}_{22}}{|\beta|^2}
        + \cdots \right]
    - \beta\left[\hat{a}_{20} + \frac{\hat{a}_{21}}{|\beta|} +
        \frac{\hat{a}_{22}}{|\beta|^2} + \cdots \right]
\nonumber\\
    &{}& - i\frac{\partial}{\partial s}\left[\hat{a}_{20} +
        \frac{\hat{a}_{21}}{|\beta|} + \frac{\hat{a}_{22}}{|\beta|^2} + \cdots \right]
    - \frac{\alpha'_2}{2} \frac{\partial^2}{\partial s^2}\left[\hat{a}_{20}
        + \frac{\hat{a}_{21}}{|\beta|} + \frac{\hat{a}_{22}}{|\beta|^2} + \cdots \right]
\nonumber\\
    &{}& + |\beta|\left(1 + \frac{i}{2\omega_0\tau_0} \frac{\partial}{\partial s}\right)
        (\hat{a}_1^2)
    = 0.
\end{eqnarray}}
Keeping terms to order $1/|\beta|$ only, we get
\begin{eqnarray} \label{eq53}
i\frac{\partial \hat{a}_1}{\partial \xi}
    - \frac{\alpha_1}{2} \frac{\partial^2 \hat{a}_1}{\partial s^2}
    + \left(1 + \frac{i}{\omega_0\tau_0} \frac{\partial}{\partial
        s}\right) (\hat{a}_1^\ast \hat{a}_{20})
    + \left(1 + \frac{i}{\omega_0\tau_0} \frac{\partial}{\partial
        s}\right) \left(\frac{\hat{a}_1^\ast \hat{a}_{21}}{|\beta|}\right)
    = 0,
\end{eqnarray}
\begin{eqnarray} \label{eq54}
i\frac{\partial \hat{a}_{20}}{\partial \xi}
    + \frac{i}{|\beta|}\frac{\partial \hat{a}_{21}}{\partial \xi}
    - \beta \hat{a}_{20}
    - \frac{\beta}{|\beta|} \hat{a}_{21}
    - \frac{\beta}{|\beta|^2} \hat{a}_{22}
    - i\frac{\partial \hat{a}_{20}}{\partial s}
    - \frac{i}{|\beta|}\frac{\partial \hat{a}_{21}}{\partial s}
\nonumber\\
    {} - \frac{\alpha'_2}{2} \frac{\partial^2 \hat{a}_{20}}{\partial s^2}
    - \frac{\alpha'_2}{2|\beta|} \frac{\partial^2 \hat{a}_{21}}{\partial s^2}
    + |\beta|\left(1 + \frac{i}{2\omega_0\tau_0} \frac{\partial}{\partial
        s}\right) (\hat{a}_1^2)
    = 0.
\end{eqnarray}

Now we collect terms of like order in (53), (54) and use
substitution to eliminate $\hat{a}_{20}$ from (54).  From terms of
lowest order (order $|\beta|$), we obtain
\begin{equation} \label{eq55}
\hat{a}_{20} = \frac{|\beta|}{\beta}\left(1 +
    \frac{i}{2\omega_0\tau_0}\frac{\partial}{\partial s}\right)
    (\hat{a}_1^2).
\end{equation}
At the next order (order $1$), we have
\begin{eqnarray} \label{eq56}
i\frac{\partial \hat{a}_1}{\partial \xi}
    - \frac{\alpha_1}{2} \frac{\partial^2 \hat{a}_1}{\partial s^2}
    + \left(1 + \frac{i}{\omega_0\tau_0} \frac{\partial}{\partial
        s}\right) (\hat{a}_1^\ast \hat{a}_{20})
    = 0,
\end{eqnarray}
\begin{eqnarray} \label{eq57}
i\frac{\partial \hat{a}_{20}}{\partial \xi}
    - \frac{\beta}{|\beta|} \hat{a}_{21}
    - i\frac{\partial \hat{a}_{20}}{\partial s}
    - \frac{\alpha'_2}{2} \frac{\partial^2 \hat{a}_{20}}{\partial s^2}
    = 0.
\end{eqnarray}
Substituting (55) into (56), the order-$1$ equation for the FF
becomes {\setlength\arraycolsep{0pt}
\begin{eqnarray} \label{eq58}
i\frac{\partial \hat{a}_1}{\partial \xi}
    &-& \frac{\alpha_1}{2} \frac{\partial^2 \hat{a}_1}{\partial s^2}
    + \frac{|\beta|}{\beta}|\hat{a}_1|^2 \hat{a}_1
    + \frac{2i}{\omega_0\tau_0}\frac{|\beta|}{\beta} |\hat{a}_1|^2
        \frac{\partial \hat{a}_1}{\partial s}
\nonumber\\
    &+& \frac{i}{\omega_0\tau_0}\frac{|\beta|}{\beta} \hat{a}_1
        \frac{\partial |\hat{a}_1|^2}{\partial s}
    - \frac{1}{\omega_0^2 \tau_0^2}\frac{|\beta|}{\beta} \left( \frac{\partial \hat{a}_1}
        {\partial s} \frac{\partial |\hat{a}_1|^2}{\partial s}
        + |\hat{a}_1|^2 \frac{\partial^2 \hat{a}_1}{\partial s^2}\right)
    = 0.
\end{eqnarray}}
Substituting (55) into (57), the order-$1$ equation for the SH
becomes {\setlength\arraycolsep{1pt}
\begin{eqnarray} \label{eq59}
2i\hat{a}_1\frac{\partial \hat{a}_1}{\partial \xi}
    -\frac{1}{\omega_0\tau_0}\left(\frac{\partial \hat{a}_1}{\partial s}
        \frac{\partial \hat{a}_1}{\partial \xi} + \hat{a}_1 \frac{\partial^2 \hat{a}_1}
        {\partial s \partial \xi}\right)
    - \hat{a}_{21}
    &-& 2i\hat{a}_1 \frac{\partial \hat{a}_1}{\partial s}
\nonumber\\
    + \left(\frac{1}{2\omega_0\tau_0} - \frac{\alpha'_2}{2}\right)
        \frac{\partial^2 \hat{a}_1^2}{\partial s^2}
    &-& \frac{i\alpha'_2}{4\omega_0\tau_0}
        \frac{\partial^3 \hat{a}_1^2}{\partial s^3}
    = 0.
\end{eqnarray}}

We can now eliminate $\frac{\partial \hat{a}_1}{\partial \xi}$ terms
by substituting (58) into (59), and the resulting expression can be
solved for $\hat{a}_{21}$.  Finally, with both $\hat{a}_{20}$ and
$\hat{a}_{21}$ known, we can write $\hat{a}_2$ to order $1/|\beta|$,
\begin{equation} \label{eq60}
\hat{a}_2 \approx \hat{a}_{20} + \frac{\hat{a}_{21}}{|\beta|}.
\end{equation}
The resulting expression, containing many terms, is then plugged
into our original equation for evolution of the FF (equation (47)),
leaving us with our desired result: an approximate single-field
expression for the FF.

We now return to our original definition of the FF, $a_1$, through
the transformation $\hat{a}_1 \rightarrow a_1/|\beta|^{(1/2)}$.  Our
single-field equation now contains terms of order $1$, $1/\beta$,
$1/\beta^2$, and $1/\beta^3$.  Of the more than 60 distinct terms in
this NLSE-like equation, only a few higher-order correction terms
are significant.  We find, to order $1/\beta^2$,
{\setlength\arraycolsep{2pt}
\begin{eqnarray} \label{eq61}
i\frac{\partial a_1}{\partial \xi}
    - \frac{\alpha_1}{2}\frac{\partial^2 a_1}{\partial s^2}
    &+& \frac{1}{\beta}\left|a_1\right|^2 a_1
    - 2i\frac{1}{\beta^2}\left|a_1\right|^2
        \frac{\partial a_1}{\partial s}
\nonumber\\
    &+& i \frac{1}{\beta} \frac{1}{\omega_0\tau_0} \left(3 \left|a_1\right|^2
        \frac{\partial a_1}{\partial s} + a_1^2\frac{\partial a_1^\ast}
        {\partial s}\right)
\nonumber\\
    &+& \left\{\textmd{h.-o. linear and nonlinear terms}\right\}
    + O\left(\frac{1}{\beta^3}\right)
    = 0,
\end{eqnarray}}
which is our approximate single-field equation for the FF.

{}

\end{document}